\documentclass[aps,twocolumn,showpacs]{revtex4}
\usepackage{graphicx}
\usepackage{dcolumn}
\usepackage{bm,amsmath,amssymb}
\begin{document}

\title{Dynamical Horizon Entropy Bound Conjecture in Loop Quantum Cosmology}
\author{Li-Fang Li}
  \affiliation{Department of Physics, Beijing Normal University, Beijing 100875, China}
\author{Jian-Yang Zhu}
\thanks{Author to whom correspondence should be addressed}
  \email{zhujy@bnu.edu.cn}
  \affiliation{Department of Physics, Beijing Normal University, Beijing 100875, China}

\begin{abstract}
The covariant entropy bound conjecture is an important hint for the
quantum gravity, with several versions available in the literature.
For cosmology, Ashtekar and Wilson-Ewing ever show the consistence
between the loop gravity theory and one version of this conjecture.
Recently, S. He and H. Zhang proposed a version for the dynamical
horizon of the universe, which validates the entropy bound
conjecture for the cosmology filled with perfect fluid in the
classical scenario when the universe is far away from the big bang
singularity. However, their conjecture breaks down near big bang
region. We examine this conjecture in the context of the loop
quantum cosmology. With the example of photon gas, this conjecture
is protected by the quantum geometry effects as expected.
\end{abstract}


\maketitle

\section{Introduction}
The thermodynamical property of spacetime is an important hint for the
quantization of gravity. Starting from Hawking's discovery of
black hole's radiation \cite{hawking75}, a
theory of thermodynamics of spacetime is being constructed gradually. Recently, the
second law of this thermodynamics was generalized to the covariant
entropy bound conjecture \cite{bousso99}. It states that the
entropy flux $S$ through any null hypersurface generated by
geodesics with non-positive expansion, emanating orthogonally from
a two-dimensional (2D) spacelike surface of area $A$, must satisfy
\begin{eqnarray}
\frac{S}{A}\leq \frac{1}{4l_p^2},
\end{eqnarray}
where $l_p=\sqrt{\hbar}$ is the Planck length. Here and in what
follows, we adopt the units $c=G=k_B=1$. Soon, Flanagan, Marolf
and Wald \cite{flanagan00} proposed a new version of the entropy
bound conjecture. If one allows the geodesics generating the null
hypersurface from a 2D spacelike surface of area $A$ to terminate
at another 2D spacelike surface of area $A'$ before coming to a
caustic, boundary or singularity of spacetime, one can replace the
above conjecture with
\begin{eqnarray}
\frac{S}{A'-A}\leq \frac{1}{4l_p^2}.
\end{eqnarray}
More recently, He and Zhang related these conjectures to dynamical
horizon and proposed a covariant entropy bound conjecture on the
cosmological dynamical horizon \cite{zhang07}: Let $A(t)$ be the
area of the cosmological dynamical horizon at cosmological time
$t$, then the entropy flux $S$ through the cosmological dynamical
horizon between time $t$ and $t'$ ($t'>t$) must satisfy
\begin{eqnarray}
\frac{S}{A(t')-A(t)}\leq \frac{1}{4l_p^2},
\end{eqnarray}
if the dominant energy condition holds for matter.

As a non-perturbative and background-independent quantization of
gravity, the loop quantum gravity (LQG) developed rapidly in recent
years. Its cosmological version, the loop quantum cosmology (LQC)
achieved many successes, including resolution of the classical
singularity \cite{bojowald01,bojowald03}, quantum suppression of
classical chaotic behavior near singularities
\cite{bojowald04a,bojowald04b}, generic phase of inflation
\cite{bojowald02,date05,xiong1} and more (for example \cite{Li}).
Since it has been suggested that the holographic principle is a
powerful hint and should be used as an essential building block for
any quantum gravity theory \cite{bousso02}, it is important and
tempting to investigate the covariant entropy bound conjecture in
the framework of the LQC, which is a successful application of the
non-perturbative quantum gravity scheme---the LQG. The authors of
\cite{ashtekar08} investigated the Bousso's covariant entropy bound
\cite{bousso99,bousso02} with a cosmology filled with photon gas and
found that the conjecture is violated near the big bang in the
classical scenario. But they found the LQC can protect this
conjecture even in the deep quantum region. In \cite{zhang07}, He
and Zhang proposed a new version of the entropy bound conjecture for
the dynamical horizon in cosmology and validated it through a
cosmology filled with adiabatic perfect fluid, governed by the
classical Einstein equation when the universe is far away from the
big bang singularity. But when the universe approaches the big bang
singularity, the strong quantum fluctuation does break down their
conjecture. In analogy to Ashtekar and Wilson-Ewing's result
\cite{ashtekar08}, one may wonder if He and Zhang's conjecture can
also be protected by the quantum geometry effect of the LQG.

Following \cite{ashtekar08}, we use photon gas as an example to
investigate this problem. As expected, we find that the loop quantum
effects can indeed protect the conjecture. Besides the result of
\cite{ashtekar08}, our result presents one more evidence for the
consistence between the loop gravity and the covariant entropy
conjecture. This paper is organized as follows. In Sec. \ref{Sec.2},
we briefly review the framework of the effective LQC and describe
the covariant entropy bound conjecture proposed by He and Zhang
\cite{zhang07}. Then in Sec. \ref{Sec.3}, we test this conjecture
with cosmology filled with photon gas, and show that the LQC is able
to protect the conjecture in all. We conclude the paper in Sec.
\ref{Sec.4} and discuss the implications.

\section{The effective framework of LQC and covariant entropy bound conjecture}
\label{Sec.2} To a good degree of approximation, the universe can
be described by the well-known Friedmann-Robertson-Walker (FRW) metric,
\begin{eqnarray}
ds^2=-dt^2+a^2(t)\left(\frac{dr^2}{1-kr^2}+r^2d\Omega^2\right),
\end{eqnarray}
which describes a homogeneous and isotropic universe. Here the
spatial curvature $k=-1,0,1$ correspond to open, flat and closed
universes respectively, and $a$ is the scale factor of the
universe. For simplicity we focus on a flat universe in this
paper. In the LQC, the phase space for a flat universe is spanned
by the coordinates $c=\gamma \dot{a}$, being the gravitational
gauge connection, and $p=a^2$, being the densitized triad.
$\gamma=0.2375$ is the Barbero-Immirzi parameter
\cite{ashtekar98}. Then the effective Hamiltonian in the LQC is
given by \cite{ashtekar06,mielczarek08}
\begin{equation}
H_{eff}=-\frac 3{8\pi \gamma ^2\bar{\mu}^2}\sqrt{p}\sin ^2\left(
\bar{\mu} c\right) +H_M. \label{hamilton}
\end{equation}
The variable $\bar{\mu}$ corresponds to the dimensionless length
of the edge of the elementary loop and is given by
\begin{equation}
\bar{\mu}=\xi p^\lambda,
\end{equation}
where $\xi>0$ and $\lambda$ depend on the particular scheme in the
holonomy corrections. In this paper we take the $\bar{\mu}$-scheme,
which gives
\begin{equation}
\xi^2=2\sqrt{3}\pi\gamma l_p^2
\end{equation}
and $\lambda=-1/2$. With this effective Hamiltonian, we have the
canonical equation
\begin{equation}
\dot{p}=\left\{ p,H_{eff}\right\} =-\frac{8\pi \gamma
}3\frac{\partial H_{eff}}{\partial c},
\end{equation}
or,
\begin{equation}
\dot{a}=\frac{\sin (\bar{\mu}c)\cos (\bar{\mu}c)}{\gamma
\bar{\mu}}.
\end{equation}
Combining with the constraint on the Hamiltonian, $H_{eff}=0$, we
obtain the modified Friedmann equation,
\begin{equation}
H^2=\frac{8\pi }3 \rho \left( 1-\frac \rho {\rho _c}\right),
\label{Friemann}
\end{equation}
in terms of the Hubble rate $H\equiv \frac{\dot{a}}a$, where $\rho :=\frac{H_M%
}{p^{3/2}}$ and $\rho _c=\frac 3{8\pi \gamma
^2\bar{\mu}^2p}=\frac{\sqrt3}{16\pi^2\gamma^2 \hbar}$. $\rho_c$ is
the critical energy density coming from the quantum effect in the
LQC. It is a large quantity and when it goes to infinity all the
quantum effects disappear.

According to \cite{zhang07}, the cosmological dynamical horizon
\cite{bousso02} is defined geometrically as a three-dimensional
hypersurface foliated by spheres, where at least one orthogonal null
congruence with vanishing expansion exists. For a sphere
characterized by any value of $(t,r)$, there are two future directed
null directions
\begin{eqnarray}
k^a_\pm=\frac{1}{a}\left(\frac{\partial}{\partial
t}\right)^a\pm\frac{1}{a^2}\left(\frac{\partial}{\partial
r}\right)^a,
\end{eqnarray}
satisfying geodesic equation $k^b\nabla_b k^a=0$. The expansion of
these null directions is
\begin{eqnarray}
\theta:=\bigtriangledown_a
k^a_\pm=\frac{2}{a^2}\left(\dot{a}\pm\frac{1}{r}\right),
\end{eqnarray}
where the dot denotes differential with respect to $t$, and the sign
$+(-)$ represents the null direction pointing to larger (smaller)
values of $r$.  For an expanding universe, i.e. $\dot{a}>0$,
$\theta=0$ determines the location of the dynamical horizon,
$r_H=1/\dot{a}$, by the definition of dynamical horizon given above.
The LQC replaces the big bang with the big bounce, so the universe
is symmetric with respect to the point of the bounce, expanding on
one side of the bounce and contracting on the other side. The
dynamical horizon in the contracting stage of the LQC corresponds to
$r_H=-1/\dot{a}$, and all of the relations are similar to the ones
given here. In this paper we only consider the expanding stage for
the LQC, but note that the contracting stage is the same.

\begin{figure}[h!]
\begin{center}
\includegraphics[width=0.45\textwidth]{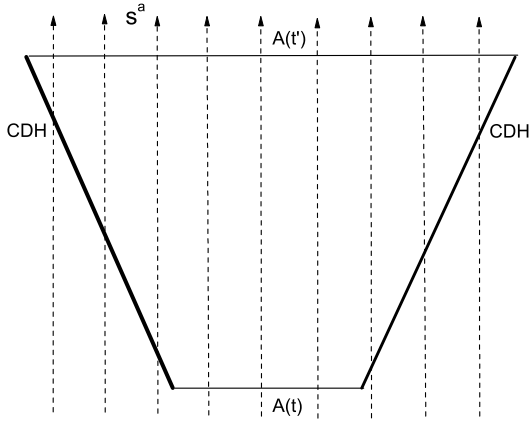}
\caption{A schematic of the entropy current flowing across the
cosmological dynamical horizon. The thick solid line marked by ``CDH''
is the cosmological dynamical horizon. The thin solid line is the
region enclosed by the CDH at time $t$ and $t'$ respectively. The
dashed lines are the entropy current.} \label{fig1}
\end{center}
\end{figure}

Since the area of the dynamical horizon is $A=4\pi a^2 r_H^2=4\pi
H^{-2}$, the covariant entropy bound conjecture in our question
becomes
\begin{eqnarray}
l_p^2S\leq\pi\left[H^{-2}(t')-H^{-2}(t)\right],
\end{eqnarray}
where $S$ is the entropy flux through the dynamical horizon
between cosmological time $t$ and $t'$ ($t'>t$), and $H$ is the
Hubble parameter. Considering that the cosmology model discussed
here is isotropic and homogeneous, we can write the entropy
current vector as
\begin{eqnarray}
s^a=\frac{s}{a^3}\left(\frac{\partial}{\partial t}\right)^a,
\end{eqnarray}
where $s$ is the ordinary comoving entropy density, independent of
space. If the entropy current of the perfect fluid is conserved,
i.e., $\nabla_a s^a=0$, $s$ will be independent of $t$ as well.
For simplicity we restrict ourselves to this special case. The
entropy flux through the dynamical horizon (shown in
Fig.\ref{fig1}) is given by
\begin{equation}
S=\int_{CDH}s^a\epsilon_{abcd}=\frac{4\pi s}{3}
\left(r^3_H(t')-r^3_H(t)\right)
\end{equation}
where $\epsilon_{abcd}=a^3r^2\sin\theta \left(dt\wedge dr\wedge
d\theta\wedge d\phi\right)_{abcd}$ is the spacetime volume 4-form.
So the conjecture is reduced to
\begin{eqnarray}
H^{-2}(t')-\frac{4}{3}l_p^2s \dot{a}^{-3}(t')\geq
H^{-2}(t)-\frac{4}{3}l_p^2s \dot{a}^{-3}(t),~~t'>t.\nonumber\\
\label{conjecture}
\end{eqnarray}

\section{Conjecture test for a cosmology filled with perfect fluid}
\label{Sec.3} Given that the FRW universe is filled with photon
gas, the energy momentum tensor can be expressed as
\begin{eqnarray}
T_{ab} &=&\rho (t)(dt)_a(dt)_b+P(t)a^2(t)\left\{ (dr)_a(dr)_b\right. \nonumber \\
&&\left. +r^2[(d\theta )_a(d\theta )_b+\sin ^2\theta (d\phi
)_a(d\phi )_b]\right\} .
\end{eqnarray}
The pressure $P$ and the energy density $\rho$ satisfy a fixed
equation of state
\begin{eqnarray}
P=\omega\rho,\label{EOS}
\end{eqnarray}
where the constant $\omega=\frac{1}{3}$. From $\nabla^aT_{ab}=0$,
we have the conservation equation
\begin{eqnarray}
\dot{\rho}+3H(\rho+P)=0.\label{continuity}
\end{eqnarray}
The comoving entropy density $s$ is given by
\begin{eqnarray}
s=a^3\frac{\rho+P}{T}=a^3(1+\omega)\frac{\rho}{T},\label{entropydensity}
\end{eqnarray}
and $\rho$ depends only on the temperature $T$,
\begin{eqnarray}
\rho=K_ol_p^{-2-\frac{1+\omega}{\omega}}T^{\frac{1+\omega}{\omega}},
\end{eqnarray}
where $K_o$ is a dimensionless constant depending on the density of
energy state of the perfect fluid. For photon gas
$K_o=\frac{\pi^2}{15}$. Plugging above thermodynamics relation into
equation (\ref{entropydensity}) we get
$s=(1+\omega)K_o^{\frac{\omega}{1+\omega}}l_p^{-1-\frac{2\omega}{1+\omega}}\rho^{\frac{1}{1+\omega}}a^3$.
Written the above conservation equation as
\begin{eqnarray}
\dot{\rho}+3(1+\omega)\rho\frac{\dot{a}}{a}=0,
\end{eqnarray}
we have an integration constant $C=\rho^{\frac{1}{1+\omega}}a^3$.
Then
$s=(1+\omega)K_o^{\frac{\omega}{1+\omega}}l_p^{-1-\frac{2\omega}{1+\omega}}C$.
 Combining our equation
of state Eq. (\ref{EOS}) with the above conservation equation, we
get the relationship between $\rho$ and the Hubble parameter,
\begin{eqnarray}
H=-\frac{1}{3(1+\omega)}\frac{\dot{\rho}}{\rho} \label{relationH}.
\end{eqnarray}
Substituting the above relation (\ref{relationH}) into the
modified Friedmann equation (\ref{Friemann}), we can get
\begin{eqnarray}
\rho=\frac{1}{6\pi(t+C_1)^2(1+\omega)^2+\frac{1}{\rho_c}}
\end{eqnarray}
where $C_1$ is an integration constant without
direct physical significance, and we can always drop it by
resetting the time coordinate. Setting $C_1=0$ gives
\begin{eqnarray}
H=\frac{ 4 \pi t  \left( 1+\omega \right)}{6\pi t^{2} \left(
1+\omega \right)
 ^{2}+\frac{1}{\rho_c}}.
\end{eqnarray}
With the definition of the Hubble parameter, we can integrate once
again to get
\begin{eqnarray}
a(t)=C^{1/3}\left[6\pi
t^2(1+\omega)^2+\frac{1}{\rho_c}\right]^{\frac{1}{3(1+\omega)}}.
\end{eqnarray}
When $\rho_c$ goes to infinity, all of the above solutions become
the same as the classical ones \footnote{Note that the original
result in \cite{zhang07} used conformal time $\eta$, while we use
universe time $t$ in this paper. $\eta$ can be negative which
divides the discussion into two cases. $t$ is always positive and
makes the discussion simpler.} presented in \cite{zhang07}. In the
classical scenario,
\begin{eqnarray}
&&H^{-2}-\frac 43l_p^2s\dot{a}^{-3}\nonumber \\
&=&\frac 94t^2(1+\omega )^2-\frac{9K_o^{\frac{\omega}{1+\omega}}l_p^{1-\frac{2\omega}{1+\omega}}}{2(6\pi)^{1/(1+\omega)}}t^{3-\frac 2{1+\omega }%
}(1+\omega )^{4-\frac 2{1+\omega }}.\nonumber \\
\end{eqnarray}
When $t\ll1$,
$H^{-2}-\frac{4}{3}l_p^2s\dot{a}^{-3}\sim-t^{3-\frac{2}{1+\omega}}=-t^{3/2}$
which is a decreasing function of $t$, so the conjecture breaks
down when the universe approaches the big bang singularity.

\begin{figure}[h!]
\includegraphics[width=0.45\textwidth]{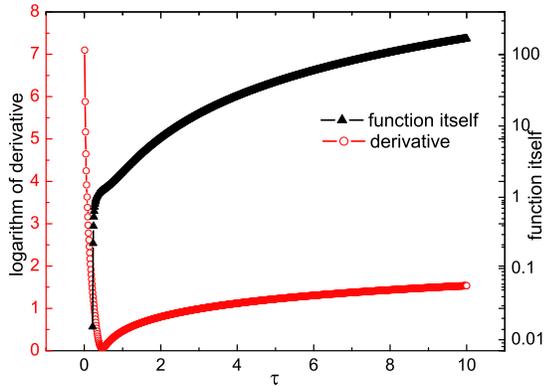}
\caption{Function $H^{-2}-\frac 43l_p^2s\dot{a}^{-3}$ and its
derivative respect to $\tau$ for photon gas.} \label{photon}
\end{figure}

We introduce a new variable $\tau=\sqrt{2\pi\rho_c}(1+\omega)t$
for the LQC to simplify the above expressions to
\begin{eqnarray}
H&=&\sqrt{2\pi\rho_c}\frac{ 2\tau}{3 \tau^{2}+1},\\
a&=&C^{1/3}\rho_c^{-\frac{1}{3(1+\omega)}}\left(3\tau^2+1\right)^{\frac{1}{3(1+\omega)}},\\
\dot{a}=aH&=&2\tau
C^{1/3}\rho_c^{-\frac{1}{3(1+\omega)}}\sqrt{2\pi\rho_c}\left(3\tau^2+1\right)^{\frac{1}{3(1+\omega)}-1}.\nonumber \\
\end{eqnarray}
Then
\begin{eqnarray}
&&H^{-2}-\frac 43l_p^2s\dot{a}^{-3}  \nonumber \\
&=&\frac 1{2\pi \rho _c}\left[ \left( \frac 32\tau +\frac 1{2\tau
}\right) ^2\right.\nonumber\\
&-&\left.\frac{(1+\omega)}{6\sqrt{2\pi}}K_o^{\frac{\omega}{1+\omega}}(\frac{\sqrt3}{16\pi^2\gamma^2})^{\frac{1}{1+\omega}-\frac12}\frac{\left(
3\tau ^2+1\right)^{3-\frac 1{(1+\omega )}}}{\tau^3}\right] . \nonumber \\
\end{eqnarray}
It is obvious that the necessary and sufficient condition for
meeting the covariant entropy bound conjecture is that the above
expression increases with $\tau$. In order to investigate the
monotone property of above function, we plot $H^{-2}-\frac
43l_p^2s\dot{a}^{-3}$ itself and its derivative respect to $\tau$
in Fig.\ref{photon}. The minimal value of the derivative is about
$1.16>0$. The covariant entropy bound conjecture for dynamical
horizon in cosmology is fully protected by loop quantum effect.

\section{conclusion and discussion}
\label{Sec.4} The covariant entropy bound conjecture comes from
the holographic principle and is an important hint for the quantum
gravity theory. In the recent years we have witnessed more and
more success of the loop quantum gravity, especially for the
problem of the big bang singularity in cosmology. The entropy
bound conjecture usually breaks down in the strong gravity region
of spacetime where the quantum fluctuation is strong, and one
would expect the loop quantum correction to protect the conjecture
from the quantum fluctuation. And Ashtekar and Wilson-Ewing do
find a result in \cite{ashtekar08} which is consistent with above
expectation. In this paper, we generalized the covariant entropy
conjecture for the cosmological dynamical horizon proposed in
\cite{zhang07} to the loop quantum cosmology scenario. We found
that the quantum geometry effects of the loop quantum gravity can
also protect the conjecture. Our result gives out one more
evidence for the consistence of covariant entropy conjecture and
loop quantum gravity theory. This adds one more encouraging result
of loop quantum gravity theory besides previous ones.

\acknowledgments The work was supported by the National Natural
Science of China (No.10875012) and the Scientific Research
Foundation of Beijing Normal University.

\end{document}